\documentclass[prl,aps,epsfig,superscriptaddress,showpacs]{revtex4}
\usepackage{bm}
\usepackage{amsfonts}
\usepackage[dvips]{graphicx}
\usepackage{mathrsfs}
\usepackage[intlimits]{amsmath}
\usepackage[colorlinks, citecolor=red]{hyperref}

\begin{document}

\title{Experimental realization of robust dynamical decoupling with bounded
controls
\\ in a solid-state spin system}
\author{F. Wang$^{1}$, C. Zu$^{1}$, L. He$^{1}$, W.-B. Wang$^{1}$, W.-G.
Zhang$^{1}$, L.-M. Duan}
\affiliation{Center for Quantum Information, IIIS, Tsinghua University, Beijing 100084,
PR China}
\affiliation{Department of Physics, University of Michigan, Ann Arbor, Michigan 48109, USA}
\date{\today }

\begin{abstract}
We experimentally demonstrate a robust dynamical decoupling protocol with
bounded controls using long soft pulses, eliminating a challenging
requirement of strong control pulses in conventional implementations. This
protocol is accomplished by designing the decoupling propagators to go
through a Eulerian cycle of the coupler group [\textit{Phys. Rev. Lett. 90,
037901(2003)}]. We demonstrate that this Eulerian decoupling scheme
increases the coherence time by two orders of magnitude in our experiment
under either dephasing or a universal noise environment.
\end{abstract}

\maketitle


\section{Introduction}

Dynamical decoupling is a powerful method to combat decoherence of quantum
systems caused by coupling to slow-varying environment \cite{1,2,3,4,5,6,7,8}%
. Conventional dynamical decoupling requires to apply a sequence of strong
instantaneous pulses \cite{1,5,8}, with the pulse duration shorter than
other time scales in the system so that the control sequence can be
approximated by $\delta $-pulses or unbounded control Hamiltonians.
Decoupling schemes based on finite-width pulses, such as the magic-echo
trains, have been proposed and implemented in experiments \cite{8a,8b},
however, they still require the driving amplitude to be much higher than
other dynamical parameters in the system \cite{8b}. Dynamical decoupling has
been used in a number of quantum information systems to suppress
decoherence, with particular applications to the solid state spin qubits
\cite{9,10,11,12,13,14,14a}. As any control Hamiltonian in real systems
always has bounded parameters, the need of strong pulses with amplitudes
larger than other dynamical parameters may impose a serious restriction on
applications of the dynamical decoupling method. Theoretically, a method
called the Eulerian decoupling scheme has been proposed to overcome this
problem \cite{6}. The Eulerian decoupling is highly robust to variation in the pulse
length and shape as long as the integral of the
pulse completes a $\pi$-rotation on the target spin, eliminating the
challenging requirement on the pulse amplitude or duration. 

In this paper, we report an experimental demonstration of effectiveness of
the Eulerian decoupling scheme.  We test performance and robustness of the 
Eulerian decoupling under various noise environments, using the solid-state
spin qubits carried by the Nitrogen-Vacancy (NV) centers in a diamond
sample. Recently, the NV centers in the diamond have stood out as a
promising system for realization of quantum information processing \cite%
{15,16,17,18,19,20,21,22,23,24,25,26}. The coherence of this system can be
well controlled and manipulated even at room temperature \cite{15,16,17}.
Here, we use this system as a testbed for various dynamical decoupling
protocols and show that the Eulerian decoupling scheme with long soft pulses
is almost equally effective as the conventional schemes based on strong
instantaneous pulses. The natural nuclear spin environment around the
diamond NV\ centers provides a source of dephasing noise. To study the
effectiveness of the Eulerian decoupling scheme under general noise
including both dephasing and relaxation, we realize a general noise model by
injecting microwave noise. Under various noise models, we demonstrate that
the corresponding Eulerian decoupling scheme can increase the coherence time
of the system by two orders of magnitude in our experiment. The
demonstration of effectiveness of the Eulerian decoupling scheme may
stimulate applications of this powerful method to systems where strong
control pulses are hard to achieve compared with the system-environment
coupling rate.

\section{Results}

\subsection{Eulerian decoupling}

In the dynamical decoupling scheme, the control Hamiltonian is denoted by ${%
H_{c}(t)}$ with its corresponding propagator given by the time-ordered
integration $U_{c}(t)=\mathscr{T}exp\{-i\int_{0}^{t}{dt^{\prime
}H_{c}(t^{\prime })}\}$. In the conventional scheme based on instantaneous
pulses, the propagator $U_{c}(t)$ suddenly changes from $g_{l-1}$ to $g_{l}$
by the pulse $l$ ($l=1,2,\cdots ,L$) with $U_{c}((l-1)t_{\Delta }+s)=g_{l-1%
\text{ }}$for $0\leq s<t_{\Delta }$, where $t_{\Delta }$ denotes the time
segment between adjacent pulses and the whole control cycle consists of $L$
pulse with total time $Lt_{\Delta }$. The evolution operators $\left\{ g_{l}%
\text{, }l=1,\cdots ,L\right\} $ form a group $G$ and the dynamical
decoupling works as we require the system-environment coupling Hamiltonian
is averaged out to zero over this group $G$. For the Eulerian decoupling
scheme, the instantaneous pulse is replaced by any continuous pulse which
satisfies the condition $U_{c}\left( lt_{\Delta }\right) =g_{l}$. For the
Eulerian decoupling scheme to work, $U_{c}\left( lt_{\Delta }\right) $ ($%
l=1,\cdots ,L$) over one complete control cycle has to follow the so-called
Eulerian path in the graph with $\left\{ g_{l}\right\} $ as the vertices. If
the noise is purely dephasing ($Z$ error), the simplest Eulerian path is
given by $\left\{ X,I\right\} $, which corresponds to the conventional
Car-Purcell-Meiboom-Gill (CPMG) echo pulse sequence $\left\{ X,X\right\} $.
We use $X,Y,Z$ in this paper to denote the three Pauli operators. So the
CPMG\ sequence should work for the dephasing noise even when we replace the
instantaneous $\pi $-pulse by a slow pulse as long as the pulse area is $\pi
$ for each control segment. If the noise is universal including both
dephasing and relaxation ($X,Y,Z$ errors), the simplest decoupling sequence
with instantaneous pulses is the XY4 sequence $\left\{ X,Y,X,Y\right\} $.
The simplest Eulerian path in this case, however, is given by $\left\{
X,-iZ,-Y,-I,-Y,-iZ,X,I\right\} $, which corresponds to the XY8 pulse
sequence $\left\{ X,Y,X,Y,Y,X,Y,X\right\} $. According to the Eulerian
decoupling scheme, the XY8 sequence should work for any slow pulse of
arbitrary shape as long as the pulse area integrated over the duration $%
t_{\Delta }$ is $\pi $ for each segment.

\subsection{Test of Eulerian decoupling under dephasing}

In our experiment, we use the electron spin qubit of a NV center in a single
crystal diamond to test effectiveness of the Eulerian decoupling scheme. Our
experimental setup is described in detail in Ref. \cite{25} and the
appendix. We compare the performance of various pulse
sequences under both dephasing and universal noise. The pulse sequences
tested in this paper are shown in Fig. 1. Note that we use the CPMG\ Y\
sequence $\left\{ Y,Y \right\} $ instead of the X sequence to combat the
dephasing noise (Z error) as the Y sequence is more robust to the systematic
X error if we prepare and measure the superposition state with X pulses in
the Ramsey measurement scheme. The dephasing noise in our experiment is
provided by the natural coupling of the electron spin to the surrounding
nuclear spins in the diamond crystal, which induces dephasing with $%
T_{2}^{\ast }\approx 1.85$ $\mu s$ measured through the free induction decay
under an external magnetic filed of $500$ G applied along the NV\ axis
(shown in Fig. 2a). To simulate an environment of universal noise, we apply
to the electron spin a noisy microwave field synthesized through an
arbitrary waveform generator (AWG) with the details shown in the
appendix.

\begin{figure}[tbp]
\includegraphics[width=85mm]{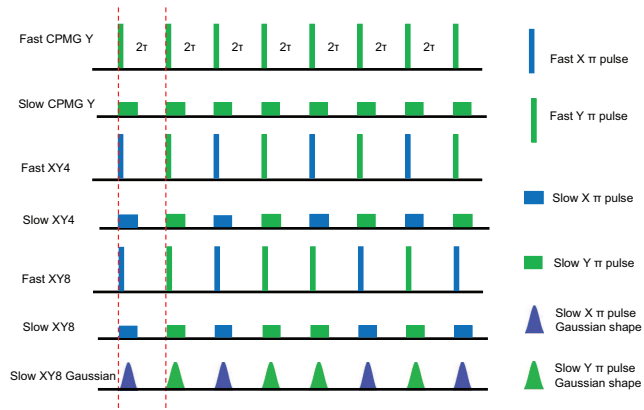}
\caption{Dynamical decoupling sequences implemented in our experiments. Each
rectangle corresponds to a $\protect\pi $ rotation of the electron spin. The
height and the width of the rectangle correspond to the strength and time
duration of the pulse, respectively. The pulse interval $2\protect\tau $ and
the pulse duration $\protect\tau_d$ together make a complete period with $%
\protect\tau_c = 2\protect\tau + \protect\tau_d$, which is kept to be the
same for different decoupling sequences in our comparison of their
performance. }
\end{figure}

We test effectiveness of the Eulerian decoupling scheme by measuring the
coherence decay of the electron spin under different dynamical decoupling
pulse sequences, with both fast and slow pulses. We prepare the initial
state in the superposition of $|0\rangle $ and $|-1\rangle $ spin state by a
$\pi /2$ pulse along the $X$ axis and then let it evolve with the decoupling
protocol for a total time duration of $N\tau_c $ (including $\tau_c /2$
before the first pulse and $\tau_c /2$ after the last pulse of the dynamical
decoupling sequence), where $N$ is the pulse number and $\tau_c $ is the
periodic pulse spacing. The performance of the dynamical decoupling is
characterized by projecting the final state $\rho $ to the $Z$ axis by
another $\pi /2$ pulse along the $-X$ axis and measuring the state fidelity $%
F=\langle 0|\rho |0\rangle $, which is calibrated by the fluorescence
contrast in a Rabi oscillation experiment.

We start by exploiting the Eulerian decoupling protocol under pure
dephasing. In Fig. 2, we show the comparison of performance under fast and
slow pulses. Note that under pure dephasing, the CMPG-Y, the XY4, and the
XY8 sequences all satisfy the Eulerian cycle condition, so theory predicts
these sequences should all work with long pulses instead of strong
instantaneous ones. In Fig. 2(b)-(d), we report experimental results with
the pulse duration extended to $500$ ns, comparable with the pulse interval $%
2\tau $ ($\tau =712$ ns) and $T_{2}^{\ast }$. Compared with the fast pulse
case (for which the pulse duration is $24$ ns), the Eulerian decoupling
sequences with slow pulses give almost the same performance. For all the
cases, the coherence time is extended to be hundreds of $T_{2}^{\ast }$, in
agreement with the prediction of the Eulerian decoupling scheme under pure
dephasing.

\begin{figure}[tbp]
\includegraphics[width=180mm, height=120mm]{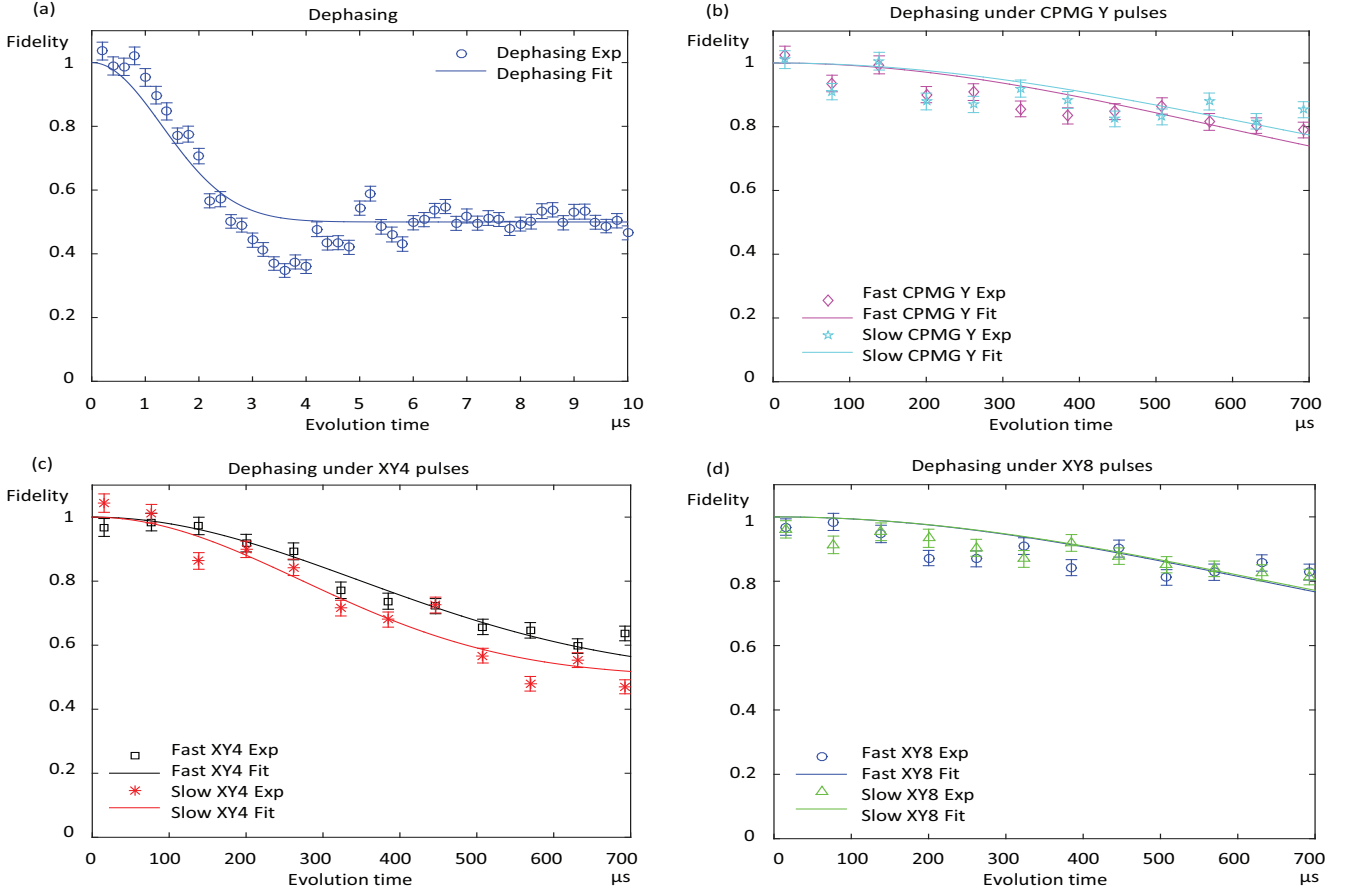}
\caption{Coherence decay with the initial states along the $+Y$ axis under
pure dephasing. (a) Free induction decay probed using Ramsey interference.
Solid line is a fit to $~exp(-(t/T_{2}^{\ast })^{2})$ with the fitted $T_{2}^{\ast }=1.85$ $\mu s$. (b,c,d) Coherence
decay under repeated CPMG Y (b), XY4 (c), and XY8 (d) dynamical decoupling
sequences compared between fast and slow
pulses, where the pulse period is fixed and the pulse number $N$ varies from $8$ to $360$ along the horizontal axis. Solid lines are fits to $~exp(-(t/T_{2})^{2})$
with (b) $T_{2}=816$ $\mu s$ ($903$ $\mu s$) for the fast (slow) CPMG pulses, (c) $T_{2}=490$ $\mu s$ ($384$ $\mu s$) for the fast (slow) XY4 pulses,
and (d) $T_{2}=884$ $\mu s$ ($896$ $\mu s$) for the fast (slow) XY8 pulses. The pulse duration $%
\protect\tau_d$ and the pulse interval $2\protect\tau $ are taken to be $24$
ns ($500$ ns) and $1900$ ns ($1424$ ns), respectively, for the fast (slow)
pulses. }
\end{figure}

\subsection{Test of Eulerian decoupling under general noise}

Universal noise provides a more interesting environment to test performance
of the Eulerian decoupling protocol. To realize a universal noise
environment, in addition to intrinsic dephasing in the diamond, we introduce
spin relaxation noise by injecting a noisy microwave field. In Fig. 3a and
3b, we characterize the universal noise environment by measuring the spin
relaxation and dephasing times under no dynamical decoupling pulses. The
magnitude of the noisy microwave filed is controlled such that the
corresponding spin relaxation time ($T_{1}^{\ast }$) is about $10$ $\mu $s\
from numerical simulation of the noise model (insert of Fig. 3a). In
experiment, however, we observe that the spin population only relax by $20\%$
after $700$ $\mu $s as shown in Fig. 3a. The observed effect can be explained by 
the dephasing induced inhabitation of relaxation \cite {25a1,25a2}. The strong
intrinsic dephasing is equivalent in role to frequent observation of population, 
which freezes population transfer by the noisy microwave field when the magnitude 
of the latter is small compared with the dephasing rate \cite {25a1,25a2}. 
Strong dephasing therefore suppresses spin relaxation. 

Although the added spin relaxation noise changes little the observed $%
T_{1}^{\ast }$ and $T_{2}^{\ast }$ in Fig. 3a and 3b, it significantly
affects the performance of the dynamical decoupling pulses. Under a
universal noise environment, theory predicts that both XY4 and XY8 sequence
should work under fast pulses, but only the XY8 sequence, which is the
minimum Eulerian decoupling sequence, will work under slow pulses. In Fig.
3c and 3d, we show the performance of the XY4 and the XY8 sequences under
both slow and fast pulses. We see that under the XY8 sequence slow pulses
achieve almost the same performance as the fast pulses, while under the XY4
sequence the performance of the slow pulses is significantly worse than that
of the fast pulses. These observations are in agreement with prediction of
the Eulerian decoupling theory.

\begin{figure}[tbp]
\includegraphics[width=180mm, height=120mm]{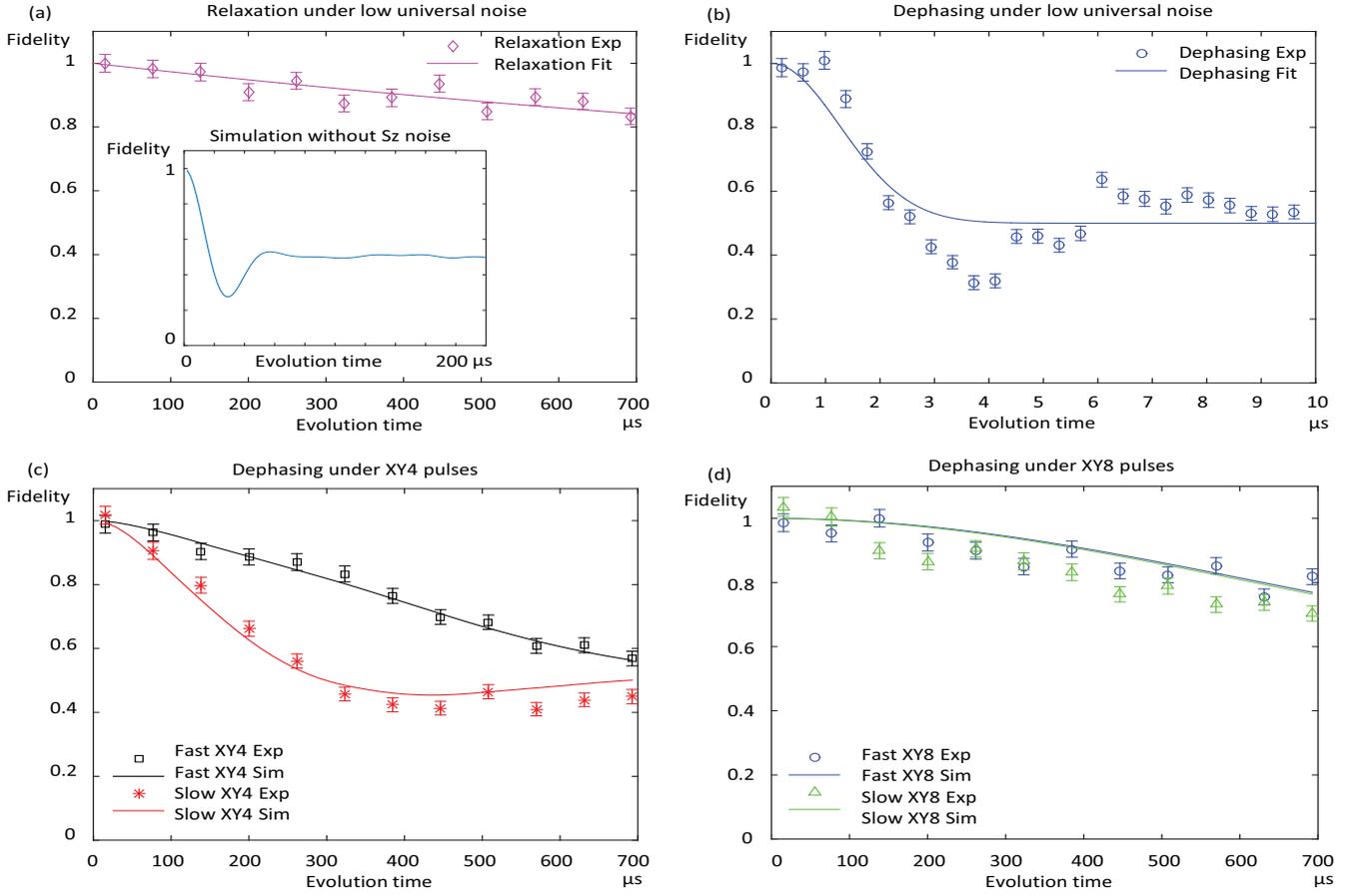}
\caption{Coherence decay with the initial states along the $+Y$ axis under
weak relaxation environment. (a) Relaxation of spin population in $|0\rangle$
state. Solid line is a fit to $~exp(-t/T_1^*)$ with the fitted $T_1^*=1.83$ ms. The inset shows the
numerical simulation result of relaxation with the injected microwave noise
if there were no dephasing in the environment, for which $T_1^*=12.87$ $\mu$s. (b) Free induction decay
measured with Ramsey interference. Solid line is a fit to $%
~exp(-(t/T_2^*)^2) $ with the fitted $T_2^*=1.79$ $\mu$s. (c,d) Coherence decay under repeated XY4 (c) and XY8
(d) dynamical decoupling sequences compared between fast and slow
pulses, where the pulse period is fixed and the pulse number $N$ varies from $8$ to $360$ along the horizontal axis. Solid lines are
numerical simulation results under the injected spin relaxation noise and the natural dephasing environment characterized in Fig. 2
(see the Appendix for the simulation method).
The pulse duration $\protect\tau_d$ and the pulse interval $2\protect\tau $
are taken to be $24$ ns ($500$ ns) and $1900$ ns ($1424$ ns), respectively,
for the fast (slow) pulses. }
\end{figure}

We then increase the spin relaxation noise by raising the magnitude of the
noisy microwave filed and test the performance of the Eulerian decoupling
scheme under strong relaxation and dephasing. The results are shown in Fig.
4. In Fig. 4a and 4b, we see that both $T_{1}^{\ast }$ and $T_{2}^{\ast }$
are now reduced to about $1$ $\mu $s. In Fig. 4c and 4d, we compare the
performance of the XY4 and XY8 sequences under slow and fast pulses with the
following two observations: First, we note that the overall coherence time
becomes shorter, and the XY4 sequence is significantly inferior to the XY8
sequence in performance even under fast pulses. This is caused by the
requirement of a smaller pulse period $\tau_c $ in this case and thus fast
accumulation of the systematic pulse errors. As it is well known, the XY8
sequence is more insensitive to accumulation of the systematic pulse errors
compared with the XY4 sequence as it suppresses the systematic errors and
the spin relaxation noise to a higher order \cite{10,14}. Under pure
dephasing, the pulse period $\tau_c $ is required to be small compared with
the bath correlation time; while under both dephasing and relaxation, the
pulse period $\tau_c $ needs to be small compared with the time scale of $%
T_{1}^{\ast }$ and $T_{2}^{\ast }$ (the inverse of the relaxation and the
dephasing rates). The latter sets a more stringent requirement and we need
to reduce $\tau_c $ by about a factor of $8$ in our experiment. The overall
coherence time correspondingly decreases under the same number of pulses.
Second, for the XY4 sequence the fast pulses significantly outperform the
slow pulses, while for the XY8 sequence they give similar results. This is
in agreement with prediction of the Eulerian decoupling theory. In Fig. 4d,
we also test performance of the slow pulses under different pulse shapes,
for instance, with a Gaussian instead of a square shape, and find that the
slow pulses under different shapes all give very similar results, with
performance comparable to that of fast instantaneous pulses.

\begin{figure}[tbp]
\includegraphics[width=180mm, height=120mm]{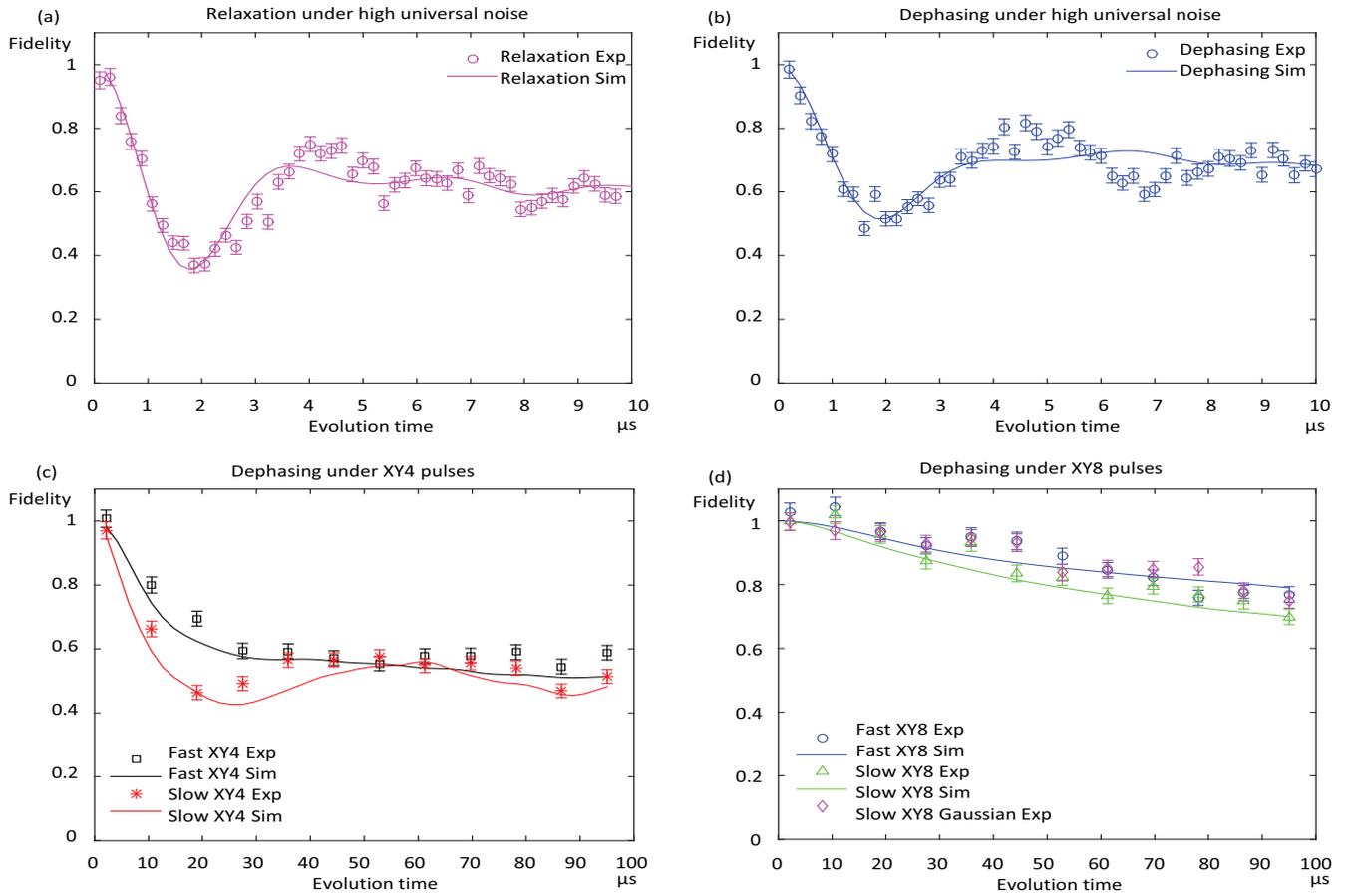}
\caption{Same as Fig.3, but under strong relaxation environment. Solid lines represent
numerical simulation results (see the Appendix for the simulation method). The pulse
duration $\protect\tau_d$ and the pulse interval $2\protect\tau $ are taken
to be $24$ ns ($100 $ ns) and $240$ ns ($164$ ns), respectively, for the
fast (slow) pulses.}
\end{figure}

\section{Summary}

In summary, we have reported an experimental test of the Eulerian dynamical
decoupling scheme in a solid state spin system and find that the slow soft
pulses, under appropriate conditions, are able to give noise suppression
comparable to the performance of strong instantaneous pulses. The
experimental demonstration of the Eulerian decoupling scheme with slow
pulses and its effectiveness eliminates a challenging requirement in
conventional dynamical decoupling techniques and may find important
applications in physical systems where the system-environment coupling is
large and it is difficult to apply strong control pulses with amplitudes
much higher than any other system or environment dynamical parameters \cite%
{8}.

\section{Appendix}

\subsection{Experimental Setup}

We use a home-built confocal microscopy to optically address single NV
centers in the diamond. A $532$ nm green laser, controlled by an acoustic
optical modulator (AOM), is used for initialization and readout of single NV
centers. The AOM is set in a double pass configuration to enhance the on-off
ratio to $100000:1$. The green laser is then coupled into a single mode
fiber for optical mode shaping. The laser coming out of the fiber is
reflected by a wave length dependent Dichroic Mirror (DM) and focused by an
oil immersed objective lens into the diamond sample, which is mounted on a $%
3 $-axis closed-loop piezo. The fluorescence of the NV center is collected
by the same objective, then passes through the DM mirror and gets collected
by a single photon detector with a $637$ nm long pass filter.

The spatial resolution of the confocal microscopy system is limited by the
diffraction limit of the green laser and the Numerical Aperture (NA) of the
objective to about $0.2$ $\mu $m. The fluorescence count of single NV center
is about $100$ k/second (the dark count rate is $5$ k/second). To collect
enough data in experiments, we repeat each experimental trial $10^{6}$
times. We apply a magnetic field of $500$ $G$ along the NV axis. Under this
filed, the green pumping laser polarizes the nearby $N^{14}$ and $C^{13}$
nuclear spins through the electron spin level anti-crossing in the
excited-state manifold, which facilitates electron-spin nuclear-spin
flip-flop process during optical pumping \cite{1}.

We fabricate a coplanar waveguide transmission line with a $70$ $\mu $m gap
on a cover glass to deliver microwave signal. The microwave signal is first
generated in a microwave source, then modulated at an IQ mixer with the
output of a frequency combiner, whose input is generated by two separate
Arbitrary Wave Generators (AWG). The first AWG generates a $100$ MHz signal
used to manipulate the electron spin (resonant with the energy level between
the $|0\rangle $ and $|-1\rangle $ states after mixed with the microwave
output). The second AWG generates the noise signal with a center frequency
same as the output of the first AWG and a particular bandwidth corresponding
to the interested noise model. The synchronization and on-off of the second
AWG is controlled by the digital output of the first AWG via a switch.

\subsection{Realization of a universal noise model by injecting microwave
noise}

The intrinsic decoherence in the diamond only provides the dephasing noise.
To model a universal noise environment, we need to add spin relaxation,
which is achieved by injecting microwave noise to drive the NV spin
transitions. The noise signal, after mixed with the microwave output, is
centered at the frequency that is resonant with the transition from the
level $|0\rangle $ to $|-1\rangle $. To model the relaxation noise, we add
up all the frequency components around the center frequency up to a cutoff
bandwidth (the bandwidth is taken to be $20$ kHz in our experiment) and
weight the frequency components with a particular spectrum. To have a time
correlation function of the shape $exp(-R|\tau |)$, we choose the weight
function to be $\sqrt{\frac{2\Delta \omega R}{(2\pi n\Delta \omega
)^{2}+R^{2}}}$ for the $n$th component of the microwave noise field with
frequency detuning $n\Delta \omega $, where ${\Delta \omega =1}$ kHz is the
discretization step. The phase of each frequency component is chosen
randomly from a uniform distribution between $0$ and $2\pi $. We generate a $%
1$ ms time trace of this spectrum and set the AWG to continuously repeat
this $1$ ms signal. All the dynamical decoupling sequence is shorter than $1$
ms, so the noise has no correlation within a cycle. We have the noise on
during the desired evolution period.

\subsection{Numerical Simulation}

Since most of the strong nuclear spins are polarized under optical pumping
with a strong magnetic field at $500$ Gauss, we ignore the interactions of
the electron spins with the nearby nuclear spins. The pulse period $\tau
_{c} $ is carefully chosen to avoid the collapse of coherence due to
resonance with the $0.5$ MHz Larmor precession induced by the external
magnetic field. We consider the effective spin-$1/2$ system composed of $%
|0\rangle $ and $|-1\rangle $ states. The numerical simulation is performed
in the rotating frame using the $4$th order Runge-Kutta integration with the
effective Hamiltonian given by $H=\Omega cos(\phi )S_{x}+\Omega sin(\phi
)S_{y}+H_{injectednoise}(t)$, where $\Omega $ and $\phi $ are respectively
the Rabi frequency and the phase of the microwave pulses for the dynamical
decoupling sequence, and $H_{injectednoise}(t)=\sum_{n=-10}^{10}W(n){cos(n\Delta
\omega t+\phi }_{n}{)S_{x}-W(n)sin(n\Delta \omega t+\phi }_{n}{)}S_{y}$ with
$W(n)=\sqrt{\frac{2\Delta \omega R}{(2\pi n\Delta \omega)^{2}+R^{2}}}$ and
random phase ${\phi }_{n}$ simulates the injected microwave noise
responsible for the spin relaxation. We perform an average over $1000$
realizations of different $H_{injectednoise}(t)$ with random phase ${\phi }%
_{n}$ in each run. The simulation result is then enveloped with the function
$0.5exp(-(t/T_{2})^{2})+0.5$ that represents the intrinsic dephasing in the
diamond, where $T_{2}$ is experimentally measured under the CPMG Y sequence.

This work was supported by the Ministry of Education of China through its
grant to Tsinghua University. LMD acknowledges in addition support from the
IARPA program, the ARL, and the AFOSR MURI program.

\end{document}